\documentclass[aps,pra,reprint,superscriptaddress]{revtex4-1}

\usepackage{amsmath,amsthm,amssymb}
\usepackage{graphicx}
\usepackage{subfigure}
\usepackage{tabularx}
\usepackage{multirow}
\usepackage{color}

\def\bra<#1|{\kern-1pt\left<#1\right|\kern-2pt}
\def\ket|#1>{\kern-1pt\left|#1\right>\kern-2pt}
\def\braket<#1>{\left<#1\right>}
\def\bracket<#1|#2>{\left<#1\vphantom{#2}\right|\left.\kern-3pt\vphantom{#1}#2\right>}
\def\braccket<#1|#2|#3>{\left<#1\vphantom{#3}\right|#2\left|\vphantom{#1}#3\right>}
\def\ketbra|#1><#2|{\left|#1\right>\kern-4pt\left<#2\right|}

\begin{document}

\title{Ancilla-Driven Universal Blind Quantum Computation}

\author{Takahiro Sueki}
\author{Takeshi Koshiba}
\affiliation{
Graduate School of Science and Engineering, Saitama University, 255
Shimo-Okubo, Sakura, Saitama 338-8570, Japan}
\author{Tomoyuki Morimae}
\affiliation{Department of Physics, Imperial College
London, London SW7 2AZ, United Kingdom}
\affiliation{ASRLD Unit, Gunma University, 1-5-1 Tenjin-cho, Kiryu-shi, Gunma 376-0052, Japan}

\begin{abstract}
Blind quantum computation is a new quantum secure protocol,
which enables Alice who does not have enough quantum technology
to delegate her computation to Bob who has a fully-fledged quantum power
without revealing her input, output and algorithm.
So far, blind quantum computation has been considered only for
the circuit model and the measurement-based model.
Here we consider the possibility and the limitation of
blind quantum computation in the ancilla-driven model, which is a
hybrid of the circuit and the measurement-based models.
\end{abstract}


\maketitle

\section{Introduction}
Traditionally, quantum computation has been studied in the circuit
model~\cite{NC00},
where the quantum register which stores quantum information consists
of many qubits, and a quantum gate operation is performed by directly
accessing one or two qubits in the quantum register.
Another canonical model of quantum computation is the one-way model~\cite{RB01}
(or more general measurement-based models~\cite{Gross,Miyake_AKLT,
Miyake_Haldane,Miyake_2dAKLT,Wei,Cai,tricluster,Raussendorf_topo,Li,FM_topo,stringnet}),
where the universal quantum
computation is performed by adaptive local measurements on a highly
entangled resource state.
Recently, a mixture of those two models, which is
called the ancilla-driven quantum computation, was proposed in \cite{AOKBA10,AABKO12}.
In this model, the quantum register is
a set of many qubits like the circuit model, whereas a
quantum gate operation is, like the one-way model, performed by
adaptive local measurements: one or two register qubits are coupled to
a single mobile ancilla, and the ancilla is measured after establishing
the interaction between the ancilla and register qubit(s).
The backaction of this measurement provides the desired
gate operation, such as a single qubit rotation or an entangling
$2$-qubit operation, on register qubit(s).
In the ancilla-driven model, the
universal quantum computation is performed with only a
single type of interaction (${\it CZ}$ or SWAP$+{\it CZ}$) between
the ancilla and register qubit(s). It is a great advantage for
experiments, since
in many experimental setups, implementing
various different types of interactions at the same time is very
difficult (such as the solid-based quantum computation).
Furthermore, the roles of the register and the information carrier are
clearly separated, and no direct action on the register is required.
Therefore, it is also useful for experimental systems where
measurements destroy quantum states, such as photonic systems.
In short, this model is a natural theoretical model
of the ``hybrid quantum computer'' where the flying ancilla mediates
interactions between static qubits (such as the chip-based quantum
computation \cite{ISM09,Devitt09}
or the hybrid system of matter and optical elements \cite{Spiller06,Loock08}.

In a future when a scalable quantum computer is realized,
the quantum computation should be done in the ``cloud" style,
since only limited number of people would have enough money
and technology to create and maintain quantum computers.
Blind quantum computation~\cite{Childs,Arrighi,Aharonov,blind_cluster,
Barz,blind_AKLT,Vedran,blind_topo,blind_measuring,FK} ensures the privacy
of the client in such a cloud quantum computing.
In protocols of blind quantum computation,
Alice, the client, does not have enough quantum technology.
On the other hand, Bob, the server, has a fully-fledged quantum power.
Alice asks Bob to perform her computation on his quantum computer
in such a way that Bob cannot learn anything about her
input, output, and algorithm.
Blind quantum computation was initially considered by using
the circuit model~\cite{Childs,Arrighi,Aharonov}.
However, in this case, Alice needs a quantum memory.
Recent new ideas of blind quantum computation which use measurement-based
models have succeeded to exempt Alice from a quantum memory~\cite{blind_cluster,Vedran,Barz,blind_AKLT,blind_topo,blind_measuring,FK}.

In terms of the computational power, 
measurement-based models do not offer any advantage
over the circuit model,
since the circuit model can be simulated by measurement-based models
and vice versa.
However, measurement-based models have provided
new points of view for studying quantum computation, and
in fact such new viewpoints have enabled plenty of successes
which have never been done in the circuit model,
such as the high-threshold 
fault-tolerancy~\cite{Raussendorf_topo,Sean,Fujii_Tokunaga,Li2,Li,
FM_topo,Nielsen_Dawson,Aliferis,Varnava,csrsnoise}, 
clarifying roles of entanglement played in quantum computation~\cite{Nest,Gross,Gross2,Bremner,ent_oneway},
and relations to condensed-matter physics~\cite{VBS,Nest2,Nest3,Gross,Miyake_AKLT,Miyake_Haldane,Miyake_2dAKLT,Wei,FM_correlation,stringnet}.
Therefore it is important
to explore the possibility of blind quantum computation
on other models than the circuit model and measurement-based models.


\section{Ancilla-driven quantum computation}
 We first define several notations for the basis and for the basic
 transformations as follows:
 $\ket|+_{\theta,\varphi}>=\cos(\frac{\theta}{2})\ket|0>+{\rm
 e}^{i\varphi}\sin(\frac{\theta}{2})\ket|1>$,
 $\ket|-_{\theta,\varphi}>=\sin(\frac{\theta}{2})\ket|0>-{\rm
 e}^{i\varphi}\cos(\frac{\theta}{2})\ket|1>$, $R_x(\theta)={\rm
 e}^{-\frac{i\theta X}{2}}$ and $R_z(\theta)={\rm e}^{-\frac{i\theta
 Z}{2}}$. We conventionally use the notations $\{\ket|\pm>\}$ and
 $\{\ket|0>,\ket|1>\}$ to denote the bases along $X$ and $Z$ axes in
 the Bloch sphere, respectively.
 Measurement outcome is represented by $s\in\{0,1\}$, associated with
 $\pm$. We denote the $i$th measurement outcome by $s_i$.

\begin{figure}[htbp]\footnotesize 
  \begin{minipage}{.49\hsize}
   \begin{center}
    \subfigure[]{\includegraphics[scale=0.35]{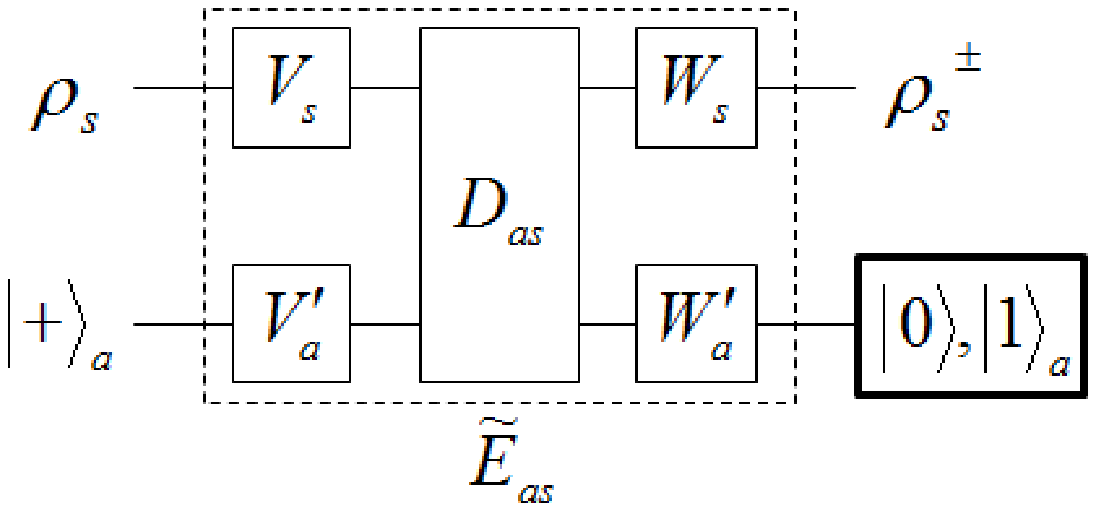}}
   \end{center}
  \end{minipage}
  \begin{minipage}{.49\hsize}
   \begin{center}
    \subfigure[]{\includegraphics[scale=0.35]{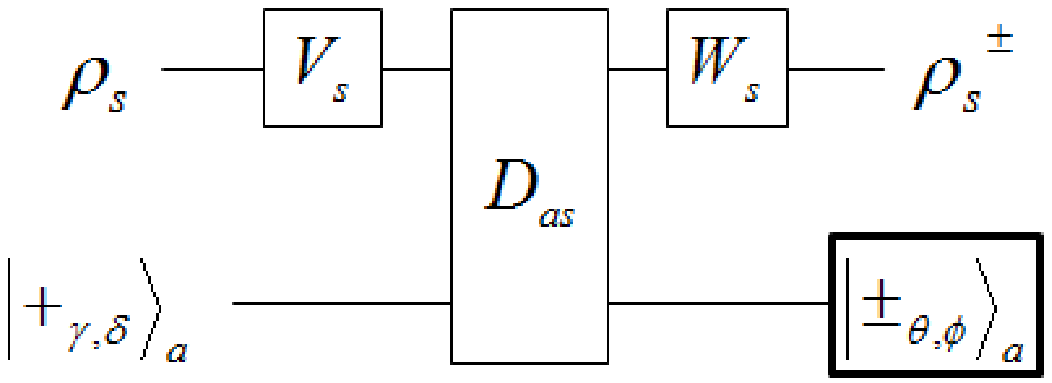}}
   \end{center}
  \end{minipage}
  \vspace{-1.2\baselineskip}
  \caption{\footnotesize ADQC. A rectangle box with bold line
  represents a measurement and the inside represents a basis for the
  measurement.}
  \label{fig:ADQC}
 \end{figure}

 We review the ancilla-driven quantum computation (ADQC) proposed
 in \cite{AOKBA10,AABKO12}. ADQC is performed with a $1$-qubit ancilla,
 only on which we can make measurements, and (a single or a few)
 $2$-qubit entangle operator(s) $\tilde{E}_{\it as}$. As in
 Fig.\ref{fig:ADQC}-(a), $\tilde{E}_{\it as}$ can be decomposed into
 $\tilde{E}_{\it as}=(W_s\otimes W'_a)D_{\it as}(V_s\otimes V'_a)$
 by the Cartan decomposition \cite{ZVSW03}, where $V_s,V'_a,W_s$ and
 $W'_a$ are $1$-qubit local unitaries and $D_{\it as}$ is a $2$-qubit non-local
 unitary. Fig.\ref{fig:ADQC}-(a) can be rewritten to
 Fig.\ref{fig:ADQC}-(b) by applying $V'_a$ to the prepared ancilla state
 $\ket|+>_a$ and $W'_a$ to the measurement basis
 $\{\ket|0>,\ket|1>\}$. $D_{\it as}$ is described as
 \begin{align*}
  D_{\it as}&={\rm e}^{-i(\alpha_xX_a\otimes X_s+\alpha_yY_a\otimes
  Y_s+\alpha_zZ_a\otimes Z_s)}
 \end{align*}
 by using non-symmetric parameters
 $0\leq\alpha_x,\alpha_y,\alpha_z\leq\frac{\pi}{4}$ due to the Weyl
 chamber \cite{Tucci05}.

 For universal quantum computation, we should choose all the parameters
 appropriately. To this end, Anders {\it et al.\/} in \cite{AABKO12}
 derive sufficient conditions for (i) Unitarity, (ii) {\em One-step\/}
 Correctable Branching, (iii) Standardization and (iv)
 Universality. Especially, we will discuss about One-step Correctable
 Branching, which states that the generalized Pauli correction according
 to the measurement outcome after ``one'' execution in ADQC enables
 the Kraus operator acting on the system deterministic. In order
 to fulfill these conditions, it is shown that the entangle operator
 $\tilde{E}_{\it as}$ must be locally equivalent to either SWAP$+{\it
 CZ}$ ($\alpha_x=\alpha_y=\frac{\pi}{4},\alpha_z=0$) or ${\it CZ}$
 ($\alpha_x=\frac{\pi}{4},\alpha_y=\alpha_z=0$). 

 \section{Ancilla-driven universal blind quantum computation}
 ADQC of SWAP$+{\it CZ}$ type can be considered as an extension of
 one-way quantum computation because the measurements are made on the
 system instead of on the ancilla if we exclude the SWAP and this case
 is exactly one-way quantum computation. So we can perform
 universal blind ADQC of SWAP$+{\it CZ}$ type as in
 \cite{blind_cluster}. In this paper, we focus only on
 universal blind ADQC of ${\it CZ}$ type. 
 Requiring all the conditions (i)-(iv) is too strong for universal
 blind
 ADQC of ${\it CZ}$ type, since ADQC of ${\it CZ}$ type satisfying all
 the conditions cannot be blind (in the sense of
 \cite{blind_cluster}) as is shown in the following. 
 The system Kraus operator for $\tilde{E}_{\it as}$
 is specified as $\tilde{K}_s^{\pm}=V_sK_s^{\pm}W_s$ and
 $K_s^{\pm}={}_a{\hspace{-2pt}}\bra<\pm_{\theta,\varphi}|D_{\it as}\ket|+_{\gamma,\delta}>_a$. 
 As in \cite{AABKO12}, Unitarity and One-step Correctable
 Branching require that the parameters for the ancilla satisfy
 $\sin\theta\cos\gamma\sin\phi=\cos\theta\sin\gamma\sin\delta$ and the Kraus
 operator $K_s^{\pm}=f_\pm I+i(-1)^{n_\pm}g_\pm X$, where $n_\pm$ are
 integers that differ in the parity. These coefficients are rewritten as
 \begin{align*}
  f_\pm&={\textstyle
  \frac{\cos\alpha_x}{\sqrt{2}}}\sqrt{1\pm\cos\gamma\cos\theta\pm\sin\gamma\sin\theta\cos(\delta-\phi)}\ 
  {\rm and} \\
  g_\pm&={\textstyle \frac{\sin\alpha_x}{\sqrt{2}}}\sqrt{1\mp\cos\gamma\cos\theta\pm\sin\gamma\sin\theta\cos(\delta+\phi)}.
 \end{align*}
 Moreover, the parameter $\alpha_x$ for $D_{\it as}$ and the parameters for
 the ancilla have the following relation:
 \begin{align*}
  \tan^2\alpha_x&=\sqrt{\frac{1-(\cos\gamma\cos\theta+\sin\gamma\sin\theta\cos(\delta-\phi))^2}{1-(\cos\gamma\cos\theta-\sin\gamma\sin\theta\cos(\delta+\phi))^2}}.
 \end{align*}
 Therefore, Unitarity and One-step Correctable Branching imply that admissible
 parameters of the ancilla are classified into the following four cases.
 \begin{table}[htbp]\footnotesize
  \begin{tabular}{|c|c|c|}
   \hline \footnotesize
   prepared ancilla & \footnotesize measurement basis & \footnotesize Kraus
   operator \\ \hline\hline
   $\gamma=0$ & $\theta=0$ & $K_s^\pm=X^sI$ \\ \hline
   $\gamma=0$ & $\theta={\rm any},\phi=0 $ & $K_s^\pm=X^sR_x(\theta)$ \\ \hline
   $\gamma=\frac{\pi}{2}$ & $\theta=0$ & $K_s^\pm=X^sX$ \\ \hline
   $\gamma,\delta={\rm any}$ & $\theta=\frac{\pi}{2},\phi=0$ &
           $K_s^\pm=X^sX$ \\ \hline
  \end{tabular}
 \end{table} \\
 For ``universal'' blind computation, a rotation operator $R_x(\theta)$
 is indispensable. Thus, we consider only the second case. In that
 case, the prepared ancilla should be fixed to $\ket|0>$ since
 $\gamma=0$. This means that we cannot use a random ancilla restricted
 on some plane in the Bloch sphere to make the
 computation blind similarly in \cite{blind_cluster}.

 For that reason, we disregard One-step Correctable
 Branching for the present and derive some more admissible parameters of
 the ancilla as follows.
 \begin{table}[htbp]\footnotesize
  \begin{tabular}{|c|c|c|}
   \hline \footnotesize
   prepared ancilla & \footnotesize measurement basis & \footnotesize Kraus
   operator \\ \hline\hline
   $\gamma={\rm any}$, & $\theta={\rm any}$, &
	   $K_s^+=\cos(\frac{\theta-\gamma}{2})I-i\sin(\frac{\theta+\gamma}{2})X$ \\
   $\delta=0$ & $\phi=0$ &
	   $K_s^-=\sin(\frac{\theta-\gamma}{2})I+i\cos(\frac{\theta+\gamma}{2})X$
	   \\ \hline
   \multirow{2}{*}{$\gamma,\delta={\rm any}$} &
       \multirow{2}{*}{$\theta=\gamma,\phi=\delta$} &
           $K_s^+=I-i\sin\gamma\cos\delta X$ \\
   & & $K_s^+=(i\sin\delta-\cos\gamma\cos\delta)X$ \\ \hline
  \end{tabular}
 \end{table} \\
 Then, we have the Kraus operators written as
 \begin{align*}
  K_s^+=R_x(\gamma)\ {\rm and}\ K_s^-=XR_x(-\gamma)
 \end{align*}
 by choosing the prepared ancilla parameters $\gamma$ be any value and
 $\delta=0$ and the measurement basis parameters $\phi,\theta=0$.
 Blind ADQC of ${\it CZ}$ type is enabled by allowing the above Kraus
 operators. For blind computation, it is sufficient that all the
 information Client sends to Server is uniformly random. This is for
 hiding the rotation of unitaries to Server in the computation and we
 show a rough sketch to incorporate this idea into ADQC. First, Client chooses
 a prepared ancilla parameter $\gamma$ randomly and Client sends ancilla
 $\ket|+_{\gamma,0}>$ or $\ket|-_{\gamma,0}>=\ket|+_{\gamma+\pi,0}>$
 with equal probabilities. Then, Server performs the Kraus operator
 $R_x((-1)^s\gamma)$ using this ancilla. At this time, the ancilla is
 $\frac{1}{2}\sum_{r\in\{0,1\}}\ketbra|+_{\gamma+r\pi,0}><+_{\gamma+r\pi,0}|=\frac{I}{2}$
 as maximally mixed state. Moreover, if we assume the input state
 $\ket|\psi>=\cos\frac{\theta'}{2}\ket|+>+{\rm
 e}^{i\varphi'}\sin\frac{\theta'}{2}\ket|->$, the output state is
 $\frac{1}{2}\sum_{r\in\{0,1\}}R_x((-1)^s(\gamma+r\pi))\ketbra|\psi><\psi|R_x^\dagger((-1)^s(\gamma+r\pi))={\scriptsize
 \left(
 \begin{array}{cc}
  \cos^2(\theta'/2) & 0 \\
  0 & \sin^2(\theta'/2)
 \end{array}
 \right)}$ so this state contains no information about $\gamma$. After
 that, Client sends a measurement basis parameter $\theta$ and Server
 performs the Kraus operator $R_x(\theta)$. Therefore, the total Kraus
 operator is $R_x(\theta+(-1)^s\gamma)$ and the total rotation of the
 Kraus operator is hiding to Server because $\gamma$ is hiding. 
 Based on this idea, we relax One-step Correctable Branching to {\em
 Multiple-step} and derive a sufficient condition which can make ADQC of
 ${\it CZ}$ type blind. {\em Multiple-step\/} means that, in ``multiple''
 executions in ADQC, each Kraus operator needs not to be
 deterministic but the whole Kraus operator must be deterministic up to
 the correction.
 
 For universal blind ADQC of ${\it CZ}$ type, two types of Kraus
 operators are necessary. One type is an {\em uncorrectable\/}
 Kraus operator which depends on an outcome of the measurement, such as
 $V_sR_x((-1)^s\gamma)W_s$, performed using a prepared ancilla parameter
 $\gamma$. The other is a {\em correctable\/} Kraus operator, such as
 $V_sR_x(\theta)W_s$ up to Pauli correction, performed with a
 measurement basis parameter $\theta$. With respect to these Kraus
 operators, we consider two conditions: {\em L-hiding\/} and {\em G-hiding}. L-hiding requires that
 $W_sR_x(\theta)V_sW_sR_x((-1)^s\gamma)V_s=W_sR_x(\theta')V_sW_sV_s\overset{\rm
 def}{=}S$ holds, where $\theta'=\theta\pm(-1)^s\gamma$. G-hiding requires
 that a gate pattern which can perform both $U\otimes U'$ where $U$ and
 $U'$ are any $1$-qubit unitaries and one kind of
 entangle operators is composable by using a unitary $S$ and a
 controlled-Pauli that can be simulated. In L-hiding, we might use an
 {\em assistant\/} Kraus operator, such as $W_sR_x(0)V_s$, for satisfying
 universality.

\begin{figure}[htbp]
  \begin{center}
   \includegraphics[scale=0.25]{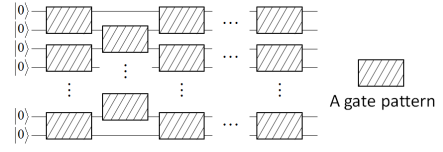}
  \end{center}
  \vspace{-1.8\baselineskip}
  \caption{\footnotesize Universal gate pattern.}
  \label{fig:pattern}
  \vspace{-1.2\baselineskip}
 \end{figure}

 If the two conditions are satisfied, we can perform universal quantum
 computation by tiling the gate pattern in G-hiding regularly as in
 Fig.\ref{fig:pattern}. What unitaries the gate pattern performs depends
 on a parameter $\theta'$ of each gate in L-hiding composing the gate
 pattern. When Client decides a parameter $\theta'$, Client sends a
 measurement basis parameter $\theta$ such that
 $\theta=\theta'\mp(-1)^s\gamma$. By choosing a prepared ancilla parameter
 $\gamma$ randomly, $\theta$ also looks random to Server. This process is
 performed similarly to the protocol in \cite{blind_cluster}. We use the
 following protocol for the performing of each $S$.
 \begin{enumerate}
  \setlength{\parskip}{-1pt}
  \item Client chooses a prepared ancilla parameter $\gamma$ randomly
	and sends the ancilla to Server.
  \item Server performs $W_sR_x((-1)^s\gamma)V_s$ with the given
	ancilla. Server sends an outcome $s$ of the measurement in this
	simulation to Client.
  \item Client decides $\theta'$ and calculates
	$\theta=\theta'\mp(-1)^s\gamma+r\pi$ with a random bit $r\in\{0,1\}$
	then sends $\theta$ to Server.
  \item Server performs $W_sR_x(\theta)V_s$ and sends an outcome $s'$ of the
	measurement in this simulation to Client.
  \item Client inverts $s'$ if $r=1$.
 \end{enumerate}
 
 If we use an assistant Kraus operator in the protocol, 
 Server performs the corresponding simulation in Step 2. 
 In the above protocol, each ancilla state is maximally
 mixed and each $\theta$ looks random to Server. Therefore, the
 information leaked to Server is only the upper bound on 
 the size of the universal gate
 pattern, that is, the upper bounds on 
 the input size and the depth of the computation. 

\begin{figure}[htbp]
  \subfigure[$R_x(\theta')$ such that
  $\theta'=-\theta+(-1)^{s_1}\gamma$]{\includegraphics[scale=0.35]{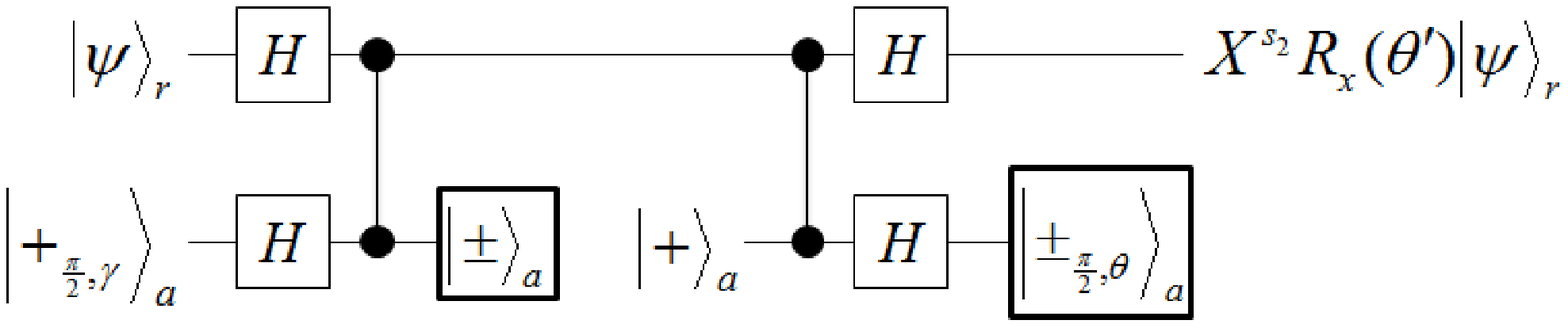}}
  \subfigure[$R_z(\theta')$ such that
  $\theta'=\theta-(-1)^{s_1}\gamma$]{\includegraphics[scale=0.35]{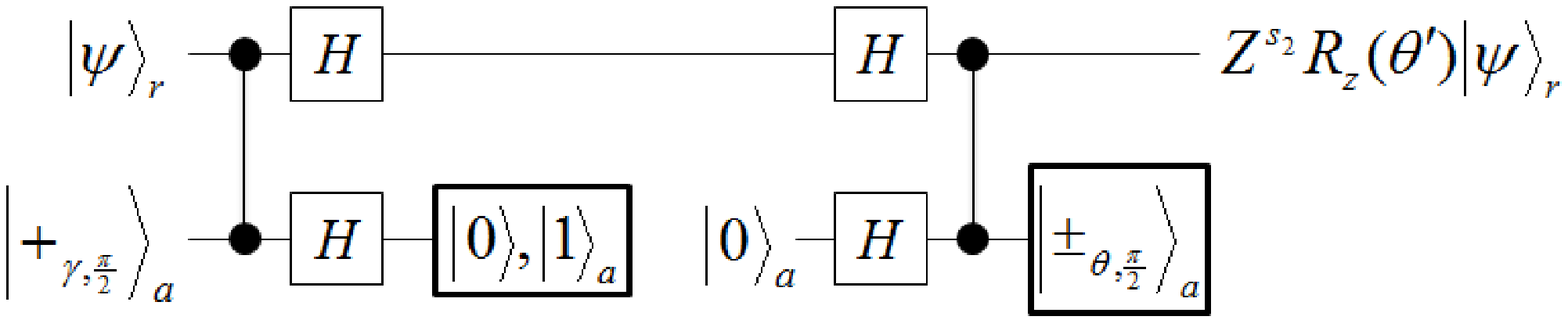}}
  \vspace{-.9\baselineskip}
  \caption{\footnotesize Simulating $R_x(\theta')$ and $R_x(\theta')$.}
  \label{fig:double}
 \end{figure}

 In the rest of this section, we discuss relations between the compatibility
of the two hiding conditions and Universality.
 In the case the protocol uses only one kind of entangle operators
 (e.g., $(H\otimes H){\it CZ}$ used in \cite{AOKBA10,AABKO12}) 
 and no assistant Kraus operator, L-hiding and G-hiding
 do not
 hold simultaneously.
 From L-hiding, it must hold that
 $V_sW_sR_x((-1)^s\gamma)=R_x(\pm(-1)^s\gamma)V_sW_s$. 
 By the similar discussion for Universality in \cite{AABKO12},
 $V_sW_s$ must be $aI+ibX$ or $aY+bZ$ up to a global phase, where
 $a,b\in {\mathbb R}$. Therefore, any $1$-qubit unitary $U$
 composed of $S$ is described as $U=W_s\tilde{U}V_s$ with the kernel
 $\tilde{U}$ which moves a quantum state only in some plane of the Bloch
 sphere, parallel to the $Y$-$Z$ plane. If $V_s$ and $W_s$ are
 determined, $U$ becomes a unitary which moves a quantum state only in
 one plane of the Bloch sphere so we cannot perform any arbitrarily rotation
 $U\otimes U'$ in G-hiding.
 \begin{figure}[htbp]
  \subfigure[{$HR_z(\theta')$ such that
  $\theta'=-\theta-(-1)^{s_1}\gamma$}]{\includegraphics[scale=0.29]{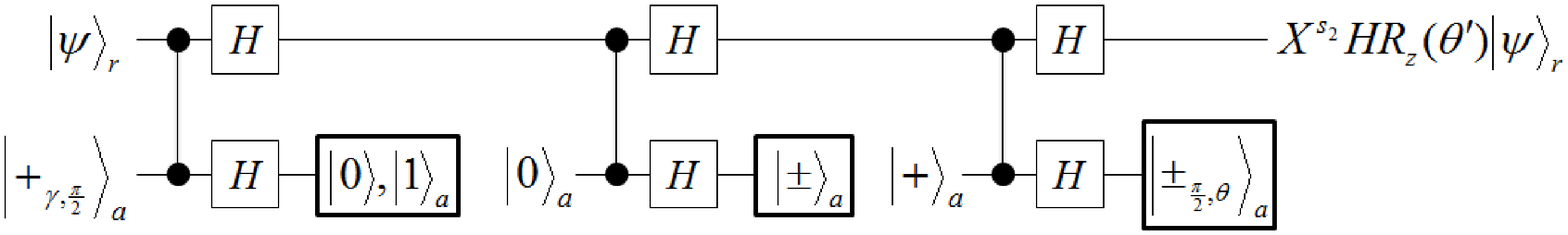}}
  \subfigure[${\it CZ}$]{\includegraphics[scale=0.29]{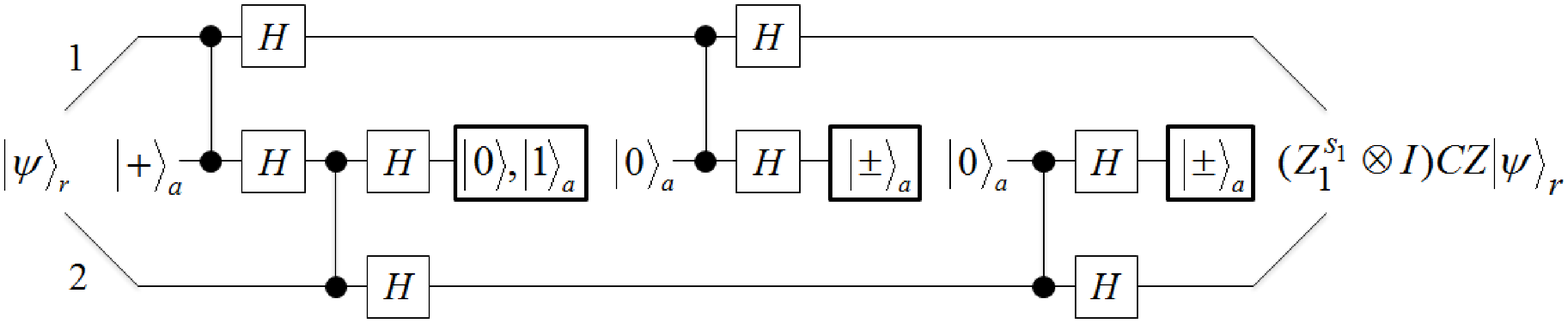}}
  \vspace{-1.4\baselineskip}
  \caption{\footnotesize Simulating $HR_z(\theta')$ and ${\it CZ}$.}
  \label{fig:single}
 \end{figure}
 \vspace{-\baselineskip}
 \begin{figure}[htbp]
  \begin{minipage}{.49\hsize}
   \begin{center}
    \subfigure[]{\includegraphics[scale=0.35]{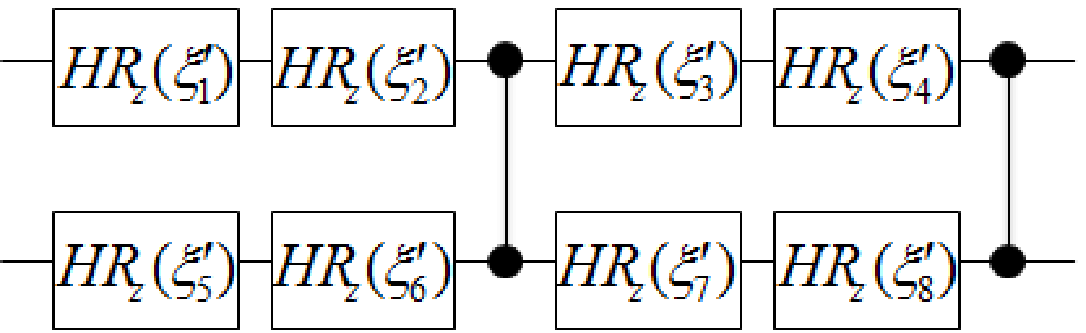}}
   \end{center}
  \end{minipage}
  \begin{minipage}{.49\hsize}
   \begin{center}
    \subfigure[]{\includegraphics[scale=0.35]{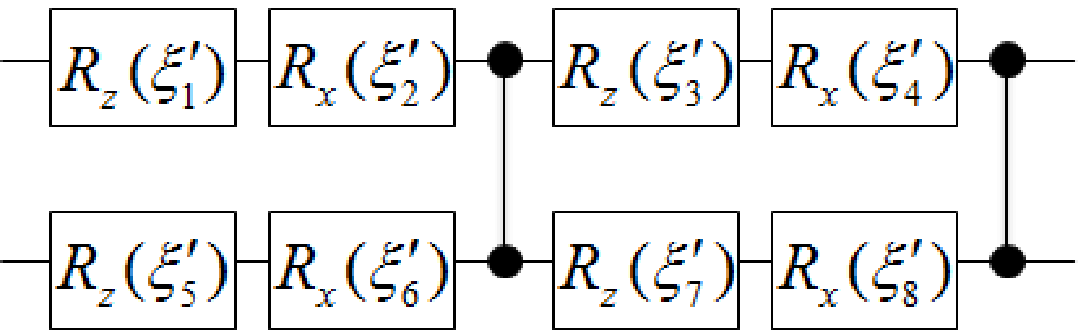}}
   \end{center}
  \end{minipage}
  \vspace{-1.4\baselineskip}
  \caption{\footnotesize Gate pattern. (a) is for a single entangle
  operator and (b) for two entangle operators.}
  \label{fig:GP}
  \vspace{-.5\baselineskip}
 \end{figure}
 In the case the protocol uses an assistant Kraus operator,
 the two hiding conditions can hold simultaneously even if
 one kind of entangle operators
 (e.g., $(H\otimes H){\it CZ}$) 
 is allowed
 to use. It is enough to show how to simulate it and the gate
 pattern. The performing in L-hiding and a controlled-Pauli (${\it CZ}$)
 in G-hiding are shown in Fig.\ref{fig:single}. The gate pattern in
 G-hiding is shown in Fig.\ref{fig:GP}-(a). 
 In the case the protocol is allowed to use two kinds of entangle
 operators, the two hiding conditions can also hold simultaneously even if
 the protocol uses no assistnat Kraus operator.
 To see that,
 the simulation in L-hiding is shown in Fig.\ref{fig:double} and the gate
 pattern in G-hiding is shown in Fig.\ref{fig:GP}-(b). In summary,
 universal blind ADQC of ${\it CZ}$ type is possible by considering
 {\em 3-step\/} Correctable Branching in the case 
 one kind of entangle operators is allowed in the protocol and
 {\em 2-step\/} Correctable Branching in the case 
 two kinds of entangle operators are allowed.

 \section{conclusion}
 In this paper, we considered the possibilities and limitations for
 universal blind computation in ADQC. First, we proved that if we
 satisfy all the conditions for universal quantum computation in
 \cite{AABKO12}, we can not perform universal blind
 computation. Therefore, we relaxed one condition and derived a
 sufficient condition for the blindness. Finally, we provided ways of
 universal blind computation in ADQC of ${\it CZ}$ type.

 Secondly, our way of universal blind ADQC needs less
 quantum requirements for Client than the way in \cite{blind_cluster} in
 the case of using quantum inputs. In our ways, Client does not need to
 rotate input states with respect to Z axis in the Bloch sphere
 but only to apply $Z$. Extending our way to one-way model, we can also
 consider a way which is also less quantum requirements in one-way model.

\end{document}